# Salt-Driven Assembly of Magnetic Silica Microbeads with Tunable Porosity

David Franck Frederic Brossault [1,2] (dffb2@cam.ac.uk) and Alexander F. Routh [1,2]* (afr10@cam.ac.uk, 01223 765718)

[1.] BP Institute, Bullard Laboratories, Madingley Rd, Cambridge CB3 0EZ, United-Kingdom
[2.] Department of Chemical Engineering and Biotechnology, Cambridge University, Philippa Fawcett Dr, Cambridge CB3 0AS, United-Kingdom

**Abstract**

Porous magnetic silica beads are promising materials for biological and environmental applications due to their enhanced adsorption and ease of recovery. This work aims to develop a new, inexpensive and environmentally friendly approach based on agglomeration of nanoparticles in aqueous droplets. The use of an emulsion as a geometrical constraint is expected to result in the formation of spherical beads with tunable composition depending on the aqueous phase content.

Magnetic silica beads are produced at room temperature by colloidal destabilization induced by addition of $CaCl_2$ to a water-in-oil emulsion containing $SiO_2$ and $Fe_3O_4$ nanoparticles. The impact of the salt concentration, emulsification method, concentration of hydrophobic surfactant as well as silica content is presented in this paper.

This method enables the production of spherical beads with diameters between 1 and 9 µm. The incorporation of magnetic nanoparticles inside the bead's structure is confirmed using Energy Dispersive X-ray spectrometry (EDX) and Scanning Transmission Electron Microscopy (STEM) and results in the production of magnetic responsive beads with a preparation yield up to 84%. By incorporating the surfactant Span 80 in the oil phase it is possible to tune the roughness and porosity of the beads.

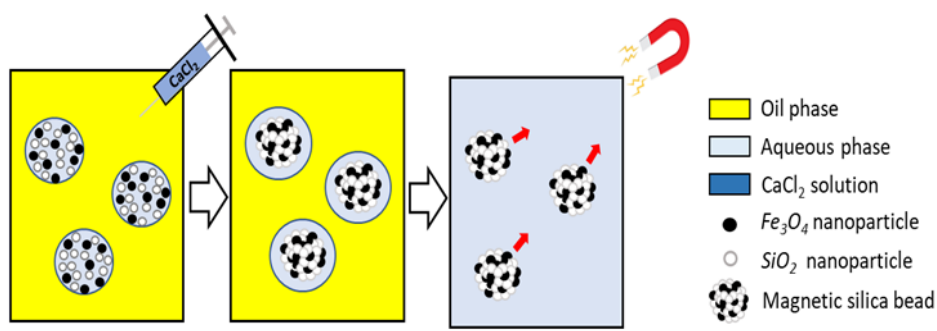

## 1. Introduction

Silica nanoparticles have attracted interest due to their low cost, insignificant toxicity, adsorption properties, surface functionalization as well as high chemical and thermal stability. [1-4] These led to such particles being considered as promising candidates for biological [3, 4] and environmental [5] applications. The usual preparation method of colloidal silica is based on the Stöber process, developed in 1968. [6] This consists of a sol-gel reaction with hydrolysis and condensation of tetraethyl-orthosilicate (TEOS) using ammonia as a catalyst. The Stöber method has been extensively investigated for the impact of concentrations and temperature to produce nanoscale silica particles with a well-controlled size and polydispersity. [3-4, 7] As a result, there are readily available commercial silica nanoparticles of specific diameters.

Complex systems, such as silica colloidosomes, [8] mesoporous silica nanoparticles [9] or mesoporous metal doped silica beads, [10-17] have recently been developed. Such systems present enhanced encapsulation, catalytic, adsorption or recovery properties depending of the additives incorporated in the silica structure. The incorporation of $TiO_2$ in silica beads enables the development of photocatalytic systems. [11] Magnetic silica beads were produced by incorporation of iron oxide within the silica structure. Such magnetic systems appear promising due to their encapsulation potential as well as adsorption and enhanced recovery properties. That is why silica composites have already been considered for applications in medical imaging, [12] drug delivery, [12-15] removal of toxins or metals from wastewater [16, 17] as well as purification of proteins and amino acids. [18] The usual method to prepare magnetic silica beads is composed of 3 steps based on the Stöber method. First, a thin layer of silica is grown on the surface of an iron oxide magnetic core via a sol-gel reaction with TEOS and ammonia.[12, 16] Zhao et al. considered the incorporation of a non-magnetic hematite core reduced to a magnetic core at a later stage using a mixture of $H_2$ and $N_2$.[13] Kim et al. investigated the use of multiple magnetic nanoparticles.[14] In the next step, a second layer of silica is grown via a sol-gel reaction in the presence of cetyltrimethylammonium bromide (CTAB) [12, 14, 16] or octadecyltrimethoxysilane (C18TMS) [13] acting as both a secondary surfactant and organic template. The organic template is then removed in the third stage leading to a mesoporous structure. This last stage is carried using various methods requiring either high temperatures or toxic chemicals. Kim et al. used a 3-hour reflux at 60°C and a pH of 1.4. [12] For Zhao et al. and Kim et al., a calcination at either 450°C or 550°C was necessary to remove the organic template. [13-14] Deng et al. investigated a milder method to remove the CTAB template. For this, the mesoporous magnetic silica beads were redispersed three times in acetone and refluxed at 80°C for 48 hours. [16] All the different approaches showed the production of mesoporous magnetic silica beads. However, these methods are limited by the toxicity of the organic template, the number of steps, the duration of the preparation method, the use of high temperatures and the production of acidic wastewater. [9]

This paper investigates the production of porous magnetic silica beads with a method based on colloidal destabilization of commercial nanoparticles dispersed in a water in oil emulsion. To the authors' knowledge, there are no previous publications using silica nanoparticles and salts to produce such systems. The impact of various experimental parameters such as the type and concentration of salt, emulsification method, surfactant concentration as well as silica and iron oxide concentrations were investigated.

## 2. Material and methods

### 2.1 Chemicals

The oil phase was composed of sunflower oil (Sainsburys) and different concentrations of sorbitan monooleate (Span 80, Sigma-Aldrich). The aqueous phase was composed of deionized water, ludox HS40 (40 wt% silica suspension in $H_2O$, diameter of 23 nm, Sigma-Aldrich), iron oxide ($Fe_3O_4$, dry powder, diameter of 50-100 nm, 97%, Alfa Aesar) and polyoxyethylene sorbitan monooleate (Tween 80, Merck). The pH was between 9 and 10. In this range, both silica and iron oxide nanoparticles are negatively charged, exhibiting a zeta potential of magnitude greater than -30 mV. Sodium chloride (NaCl, Sigma-Aldrich) and calcium chloride ($CaCl_2.2H_2O$, Sigma-Aldrich) were used to induce a colloidal instability. Absolute ethanol (98% v/v, Sigma-Aldrich) and polyoxyethylenesorbitan monolaurate (Tween 20, Aqueous organics) were used during the cleaning and redispersion stages. All the chemicals were employed as received without additional purification.

## 2.2 Preparation of silica beads

### 2.2.1 Overview of the preparation method

The preparation of silica beads can be divided into three stages as described in **Figure 1**: (1) emulsification, (2) colloidal destabilization and (3) cleaning and redispersion. During the first stage, an aqueous solution (1 vol %) containing silica particles (30 wt %) was added to sunflower oil and emulsified. A salt solution (1 vol %) was then added to the system, whilst stirring, enabling the nanoparticles to agglomerate forming the silica beads. The beads were cleaned and transferred to an aqueous phase. For this, the samples were first centrifuged at 3000 RPM for 10 min using a Multifuge 1S-R (Heraeus). The oil phase was then removed, and the beads cleaned with absolute ethanol (5 mL) before being redispersed into a Tween 20 solution (1 wt%, 5mL). The samples were then centrifuged a second time at 500 RPM for 5 min to remove residual oil before collecting the water dispersed beads using a Luer lock 5 mL syringe equipped with a long STERICAN needle (0.80 x 120 mm, Braun).

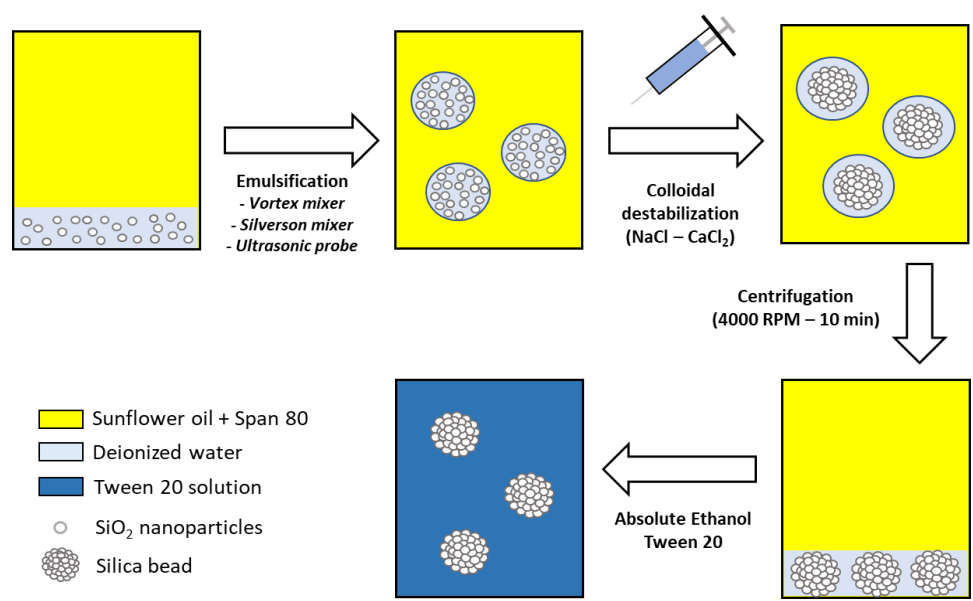

**Figure 1:** Overview of the preparation method of silica beads.

### 2.2.2 Impact of the type and concentration of salt

Two salts were investigated: one with a monovalent cation (NaCl) and one with a divalent cation ($CaCl_2$). The samples were prepared as described above following the conditions presented in **Table 1**.

### 2.2.3 Impact of the emulsification conditions on the diameter of the silica beads

Three different emulsification methods were employed: vortex mixer (TopMix FB15024, Fisher Scientific), Silverson mixer (Silverson SL2 equipped with a 4-blade radial flow rotor and a general-purpose disintegrating head, Silverson®) and ultrasonic homogenizer (Sonic Dismembrator FB-120 fitted with a Model CL-18 probe, Fisher Scientific). The samples were prepared following the conditions presented in **Table 1**.

### 2.2.4 Modification of the bead's porosity

A correlation between the concentration of Span 80 in the oil phase and the porosity of the silica beads was investigated. The formulations tested are presented in **Table 1**.

### 2.2.5 Production of silica beads with reduced solids content in the aqueous phase

Two series of experiments have been carried out. In the first, the concentration of $CaCl_2$ was kept constant at 1.25 mol L$^{-1}$. In the second series a fixed ratio between the concentration of silica and $CaCl_2$ was investigated. The samples are described in **Table 1**.

**Table 1.** Summary of the experimental conditions followed to produce silica beads.

| Sample composition | | | Emulsification conditions | | | Locking conditions [a] | | |
|---|---|---|---|---|---|---|---|---|
| Water/Oil volume [mL] | Silica [wt %] | Span 80 [wt%] | Equipment | Duration [s] | Speed [RPM] | Salt type | Salt [mol L$^{-1}$] | Volume [mL] |
| **Experimental section 2.2.2**: Impact of the type and concentration of salt | | | | | | | | |
| 1 / 99 | 30 | 0 | Silverson | 60 | 3000 | NaCl | 0, 0.1, 0.5, 1 or 1.25 | 1 |
| 1 / 99 | 30 | 0 | Silverson | 60 | 3000 | $CaCl_2$ | 0, 0.1, 0.5, 1 or 1.25 | 1 |
| **Experimental section 2.2.3**: Impact of the emulsification conditions on silica bead's diameter | | | | | | | | |
| 0.2 / 19.8 | 30 | 0 or 1 | Vortex | 10, 20, 30 or 60 | 3000 | $CaCl_2$ | 1.25 | 0.2 |
| 1 / 99 | 30 | 0 or 1 | Silverson | 10, 20 or 60 | 3000, 4000, 5000 or 6000 | $CaCl_2$ | 1.25 | 1 |
| 0.2 / 19.8 | 30 | 0 or 1 | Ultrasound | 15, 30 or 60 | Amplitude 90% | $CaCl_2$ | 1.25 | 0.2 |
| **Experimental section 2.2.4**: Modification of the silica bead's porosity | | | | | | | | |
| 0.2 / 19.8 | 30 | 0, 0.5, 1 or 2 | Vortex | 60 | 3000 | $CaCl_2$ | 1.25 | 0.2 |
| 1 / 99 | 30 | 0, 0.5, 1 or 2 | Silverson | 60 | 4000 | $CaCl_2$ | 1.25 | 1 |
| 0.2 / 19.8 | 30 | 0, 0.5, 1 or 2 | Vortex | 60 | 3000 | $CaCl_2$ | 1.25 | 0.2 |
| **Experimental section 2.2.5**: Production of silica beads with reduced solid content | | | | | | | | |
| 1 / 99 | 10, 20, 30 or 40 | 0 or 1 | Silverson | 60 | 4000 | $CaCl_2$ | 1.25 | 1 |
| 1 / 99 | 10, 15, 20, 25 or 30 | 0 or 1 | Silverson | 60 | 4000 | $CaCl_2$ | 0.42, 0.63, 0.84, 1.05 or 1.25 | 1 |

a) (Mixing for 10 seconds at 3000 RPM)

## 2.3 Preparation of magnetic silica beads

100 mL samples were prepared by mixing the aqueous phase (1 mL) with the oil phase (99 mL). The aqueous phase was composed of $SiO_2$ nanoparticles (30 wt%), Tween 80 (1 wt%) and $Fe_3O_4$ nanoparticles (0, 3, 6, 10 or 15 wt%). The oil phase was composed of sunflower oil and Span 80 (0, 1 or 2 wt%.) All the samples were emulsified using the Silverson mixer at 4000 RPM for 60 seconds. The samples were then destabilized by addition of a $CaCl_2$ solution (1mL, 1.25 mol L$^{-1}$) and stirred for 10 seconds at 3000 RPM. The solutions were split into tubes of 20 mL and centrifuged at 3000 RPM for 10 minutes. The oil phase was then removed and the beads were cleaned with absolute ethanol (5 mL) and redispersed in an aqueous solution of Tween 20 (5 mL, 1 wt %). An extra cleaning stage was applied to the system to remove any non-magnetic silica beads. For this, the magnetic beads were separated from the solution using a Rare-earth neodymium magnet (NiCuNi, 20 x 20 mm, E-magnets UK) and cleaned three times with a Tween 20 solution (1 wt%) (**Figure 2**).

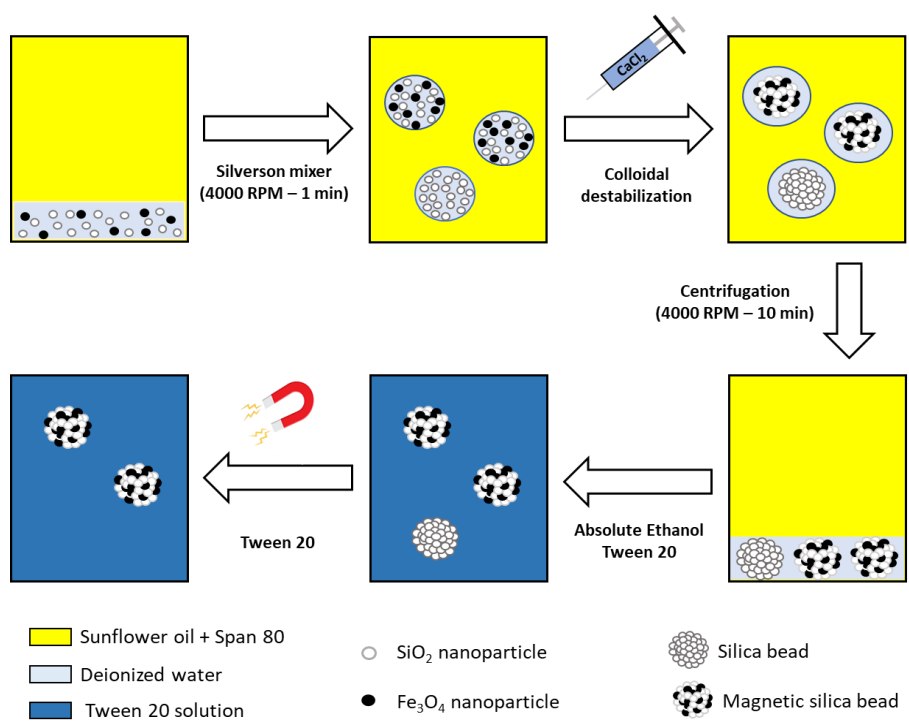

**Figure 2:** Overview of the preparation method of magnetic silica beads.

It is likely possible to produce pure magnetic beads, without silica. However, this was not investigated since the production of novel composites was aimed.

## 2.4 Characterization methods

An optical microscope (Leica DME) equipped with x 10 and x 63 magnification lenses was used to provide an overview of both the size and geometry of the beads.

For scanning electron microscopy, the samples were first dried on a cover glass slide before being coated with 10 nm of platinum using a turbomolecular pump coater (Q150T ES, Quorum). The samples were then observed at a 5-kV acceleration voltage using a Tescan Mira3 FEG-SEM. ImageJ software was used to analyze 50 randomly chosen beads and obtain an average diameter and polydispersity. The Tescan Mira3 microscope was also used to determine the samples composition via Energy Dispersive X-ray spectroscopy (EDX) using a 30-kV acceleration voltage.

Scanning transmission electron microscopy (STEM) images were taken to confirm the incorporation of $Fe_3O_4$ nanoparticles into the structure and to investigate the porosity. For this, 3 µL of solution was applied on a TEM membrane (Carbon filmed copper grid, Agar Scientific). The samples were dried, plasma cleaned for 30 seconds (Model 1070 Nanoclean, Fischione instruments) and observed using a Talos F200X G2 from thermo scientific equipped with a Ceta 16M camera at an acceleration voltage of 200 kV.

Porosity characterizations were carried out using BET measurements (TriStar II Plus, Micromeritics). For this, magnetic silica beads were dried at 100°C for 4 hours in a vacuum oven (Gallenkamp). Between 150 and 300 mg of sample was held for 8 hours at 120 °C under vacuum before measurement.

An estimation of the bead stability was obtained by zeta potential measurements using a Brookhaven ZetaPALS. The quoted value was the average of 5 runs each consisting of 20 cycles.

The concentration and preparation yield of the magnetic silica beads were determined by drying. For this, 2 vials containing 250 µL of solution were dried in a vacuum oven at 100°C for 4 hours. The samples were then weighed, and the ratio of the mass obtained after drying to the mass of nanoparticles originally dispersed was taken as the yield. The mass of residual calcium chloride was neglected from the calculations due to the extensive cleaning of the beads.

The magnetic response of the iron oxide silica beads was assessed by observing the separation of the beads from solution under an external magnetic field. For this, 5 mL of solution containing magnetic silica beads was placed in a 7 mL glass vial and exposed to a magnetic field generated by a Rare-earth neodymium magnet (NiCuNi, 20x20mm, E-magnets UK).

## 3. Results and discussion

Addition of electrolyte to a silica dispersion results in a spatially inhomogeneous agglomerated fractal structure. [19] However, we observe the formation of spherical beads. This results from the emulsion which acts as a template and therefore sets a maximum size of the agglomerate. It is also important to highlight that the system does not form a Pickering emulsion. The silica nanoparticles are too hydrophilic to stabilize the oil/water interface. This aspect was confirmed experimentally by preparation of emulsions containing sunflower oil, water, and silica nanoparticles. Samples prepared in the presence or absence of silica nanoparticles separated into aqueous and oil phases at the same rate. This result is consistent with the water contact angle being below 20° for unmodified silica particles. [20]

### 3.1 Preparation of silica beads

*3.1.1. Impact of the type and concentration of salt*

Samples prepared using 0.1 and 0.5 mol L$^{-1}$ of NaCl resulted in the absence of silica beads. This demonstrates that a minimum salt concentration is required to destabilize silica particles. Stable systems with diameters around 2.5 ± 1.0 µm were obtained using 1 and 1.25 mol L$^{-1}$ of NaCl (**Figure 3**). However, a slight merging of the beads was observed with 1 mol L$^{-1}$. The experiments carried out using CaCl$_2$ resulted in the formation of large quantities of silica beads with a diameter around 3.1 ± 1.2 µm for salt concentrations as low as 0.1 mol L$^{-1}$ (**Figure 3**).

These experiments demonstrate the method of producing silica beads using either NaCl or CaCl$_2$. However, CaCl$_2$ appears more promising due to its ability to produce silica beads at lower concentrations.

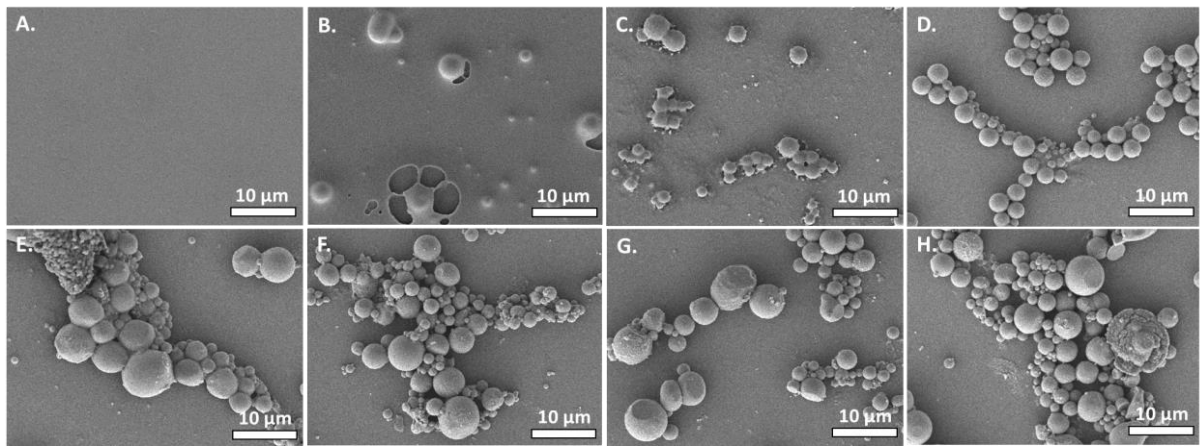

**Figure 3:** SEM images of silica beads produced using a Silverson mixer at 3000 RPM for 1 min with (**A**) 0.1 mol L$^{-1}$ NaCl, (**B**) 0.5 mol L$^{-1}$ NaCl, (**C**) 1 mol L$^{-1}$ NaCl, (**D**) 1.25 mol L$^{-1}$ NaCl, (**E**) 0.1 mol L$^{-1}$ CaCl$_2$, (**F**) 0.5 mol L$^{-1}$ CaCl$_2$, (**G**) 1 mol L$^{-1}$ CaCl$_2$ and (**H**) 1.25 mol L$^{-1}$ CaCl$_2$.

*3.1.2. Impact of the emulsification conditions on the diameter of the silica beads*

Microbeads were produced with diameters between 1.4 and 8 µm depending on the emulsification parameters. The emulsion acts as a template for the agglomeration of the silica nanoparticles, controlling the size. Using a higher energy emulsification method reduces both the emulsion droplet size and polydispersity. All the results presented in **Figure 4** were obtained by analyzing optical microscopy pictures (**Figure S1**). Those results were also confirmed in the SEM images presented in **Figure S2**. These confirmed a bead structure as opposed to the core-shell structure which would be obtained for Pickering emulsions.

Samples prepared using the vortex mixer showed the largest sizes and polydispersity with diameters between 5 and 8 µm and standard deviations up to 5.1 µm from 10 seconds of mixing. Optical microscopy pictures also highlighted a non-spherical structure of the silica beads. The vortex mixer appears an easy method to prepare silica beads. However, limits were observed regarding the volume of sample and diameter of the beads which can be produced without addition of surfactant.

Samples prepared with the Silverson mixer showed a large span of size and polydispersity depending of the speed and mixing duration. Silica beads with diameters between 1.5 and 5.0 µm were produced. An increased mixing speed gave a reduction of the emulsion droplet size, generating smaller beads. The Silverson mixer appears a good method to produce various sizes of silica beads. However, the method requires larger volumes and longer cleaning steps between samples to avoid cross-contamination.

Samples prepared with an ultrasonic homogenizer showed smaller sizes, with only 15 seconds of mixing. This high energy emulsification method appears to be an efficient method to produce silica beads with diameters between 1.4 and 2.2 µm. Submicron beads were also prepared. To achieve this, smaller volumes were produced in narrow falcon tubes to maximize the contact between the ultrasonic probe and the solution. Diameters as small as 300 nm were obtained.

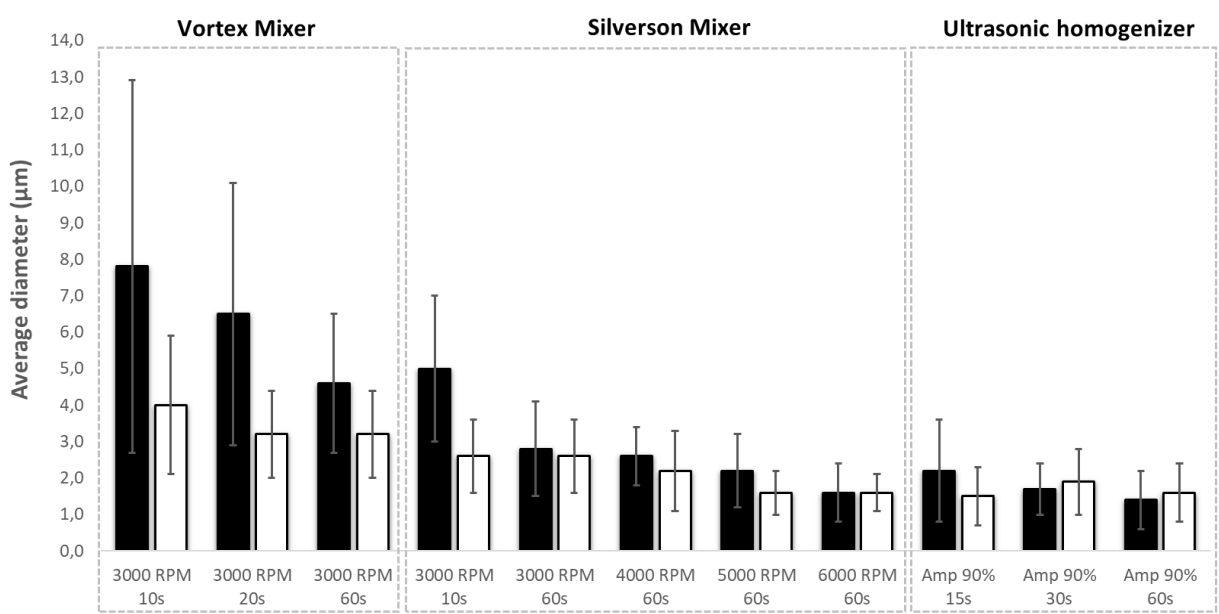

**Figure 4:** Impact of the emulsification conditions on the average diameter and standard deviation of silica beads. Samples prepared in the absence (black) or presence (white) of 1 wt% of Span 80 in the oil phase. Values based on the 50 beads randomly chosen in each sample and analyzed using ImageJ.

Adding Span 80, a hydrophobic surfactant, to the oil phase resulted in the production of smaller silica beads. The optical and electron microscopy images also revealed an impact of Span 80 on the morphology and porosity of the beads. This aspect is investigated further in the following section.

*3.1.3. Modification of the silica bead's porosity by addition of Span 80*

Samples were prepared using a vortex mixer set at 3000 RPM for 1 min and a Silverson mixer set at 4000 RPM for 1 min with various concentrations of Span 80. Similar trends were observed for both emulsification methods as shown in **Figure 5** and **Figure S3**. A variation of the bead diameter was observed by increasing the Span 80 concentration up to 2 wt%. At low concentrations, Span 80 adsorbs to the water / oil interface enabling newly generated interfaces to be stabilized, resulting in smaller water droplets and hence smaller silica beads. At higher concentrations, larger beads were observed. The reason for this is not fully understood and will be investigated further in future work. Optical microscopy and SEM images highlighted a significant change of morphology and porosity of the silica beads for increased surfactant concentrations. In the absence of Span 80, the silica beads were spherical and presented a smooth surface. Images at higher magnification confirmed a slight roughness caused by the silica nanoparticles forming the beads. An increase of the roughness was observed for the samples prepared with 0.5 wt% Span 80. That roughness expanded at higher concentration of Span 80 leading to the formation of a few pores at 1 wt% and to a highly porous structure at 2 wt%.

In the absence of Span 80, a spherical bead is observed due to the emulsion template. As discussed previously, the emulsion generates a restricted volume limiting the motion of the silica nanoparticles. The colloidal destabilization and agglomeration occur in a spherical water droplet rich in silica nanoparticles, which results in a packed spherical bead with low porosity as shown in **Figure 5**. In the presence of Span 80, there is adsorption of surfactant onto the silica nanoparticles. [21] Two potential mechanisms are considered to explain the observed change of morphology. First, the presence of Span on the surface of the nanoparticles could lead to adsorption of oil during the preparation method, resulting in the formation of a system composed of both oil and agglomerated silica nanoparticles. When the oil is washed away voids are generated in the beads and thus porous structures. Alternatively, micellar structures could also be formed at the surface of the silica nanoparticles, resulting in a steric hindrance restricting the packing of the particles. Those two hypotheses are presented in **Figure 5**.

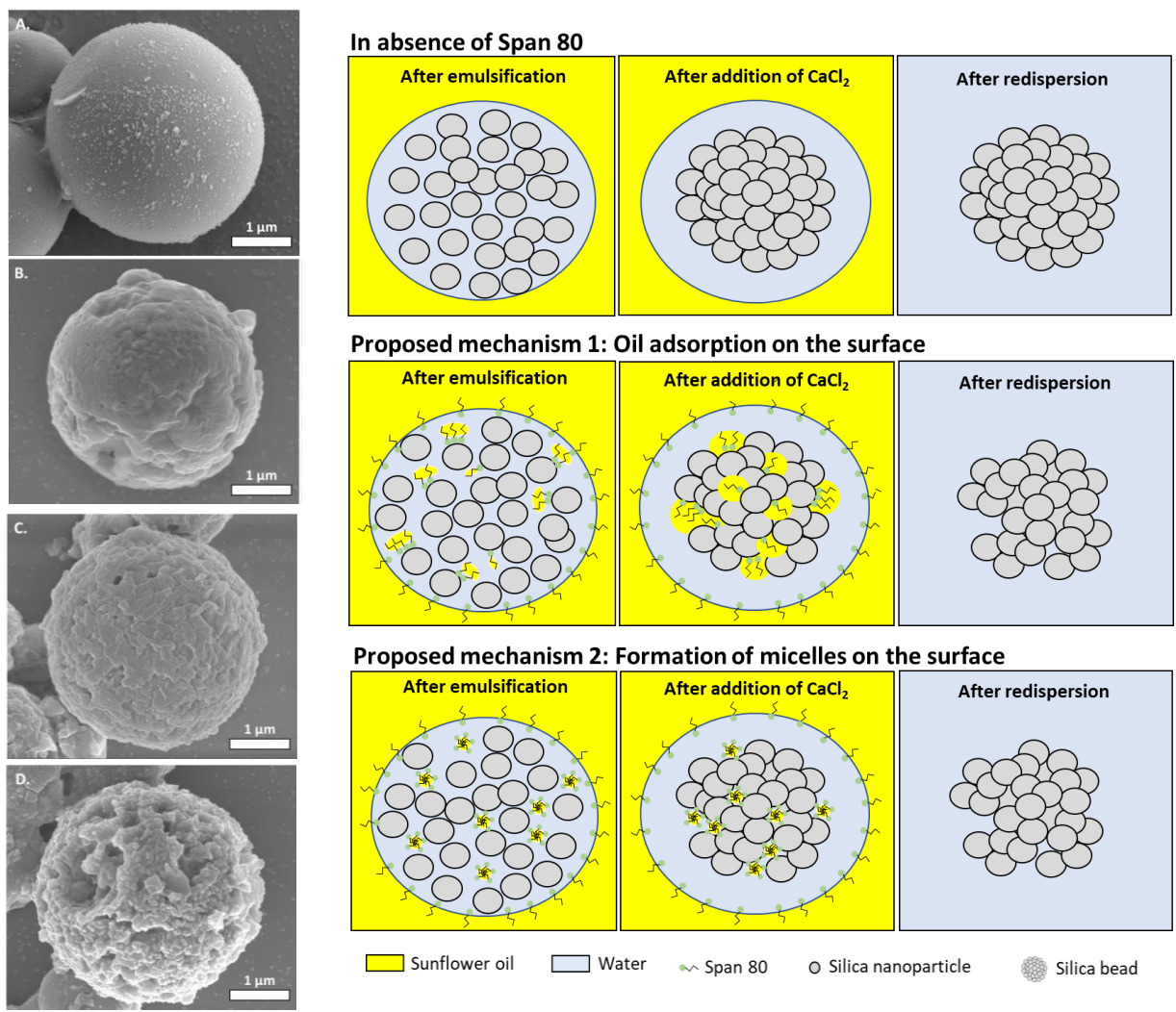

**Figure 5:** SEM images of silica beads prepared using the vortex mixer at 3000 RPM for 1 min with (**A**) 0 wt%, (**B**) 0.5 wt%, (**C**) 1 wt% or (**D**) 2 wt% of Span 80 in the oil phase. Proposed mechanisms to explain the difference of porosity observed in the presence of Span 80.

*3.1.4. Reduction of the silica content*

Samples were prepared maintaining the concentration of $CaCl_2$ at 1.25 mol $L^{-1}$. For concentrations of silica of 30 wt% and 40 wt%, SEM images (**Figure 6a**) showed the presence of spherical silica beads both with or without Span 80. In the absence of Span, spherical beads were still visible for lower concentrations of silica. However, agglomeration could also be noticed likely due to the high salt content. In the presence of Span, an agglomeration was observed for the sample prepared with 20 wt% silica nanoparticles. With 10 wt% of silica no beads were visible. This suggests that the low amount of silica and increased porosity of such structures do not provide enough strength to resist the cleaning and transfer stages.

Samples were produced using 10, 15, 20 and 25 wt% silica nanoparticles in the aqueous phase and a fixed salt / silica ratio of 0.042 mol L$^{-1}$ of CaCl$_2$ / wt% of Ludox H40 in the aqueous droplets. SEM images (**Figure 6b**) taken in the absence of Span 80 showed that the reduction of the salt concentration minimized the agglomeration of the beads, which could be produced at concentrations as low as 15 wt%. However, no capsules were obtained with 10 wt% silica and broken ones were obtained for 15 and 20 wt%. Those results suggest that a higher silica content strengthens the silica beads enabling them to resist the later preparation stages. With Span, the results are similar with broken structures visible for concentrations of silica below 20 wt%.

a)

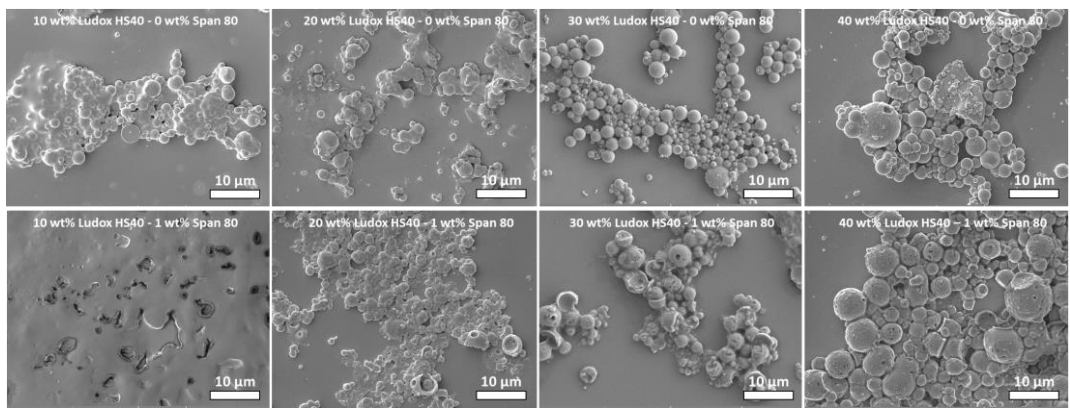

b)

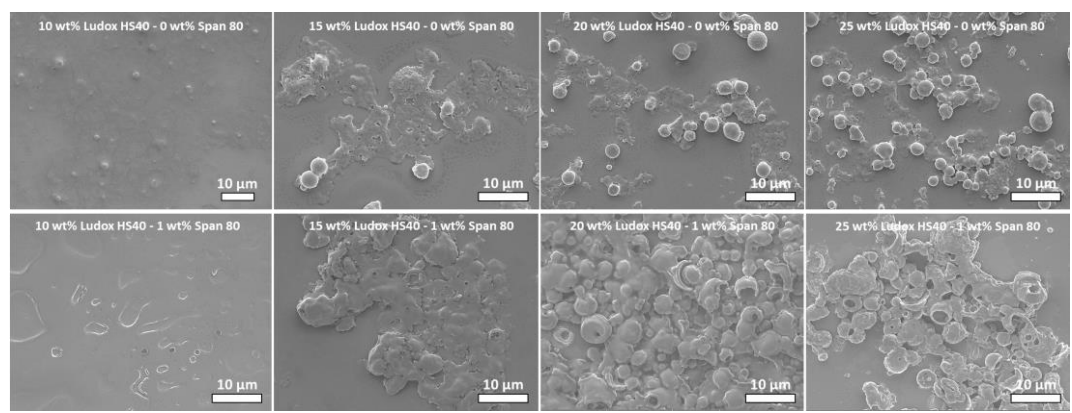

**Figure 6:** SEM images of silica beads prepared using the Silverson mixer at 4000 RPM for 60 seconds. a) Colloidal destabilization carried out using a CaCl$_2$ solution at a fixed concentration of 1.25 mol L$^{-1}$. Aqueous phase composed of 10, 20, 30 or 40 wt% of Ludox HS40. Oil phase composed of 0 or 1 wt% of Span 80. b) Colloidal destabilization carried out using with various concentrations of CaCl$_2$ solutions to maintain a constant salt / silica concentration ratio. Aqueous phase composed of 10, 15, 20 or 25 wt% of Ludox HS40. Oil phase composed of 0 or 1 wt% of Span 80.

Those experiments demonstrate that silica beads can be produced at lower silica concentrations. However, the resulting beads are weaker.

## 3.2 Production and characterization of magnetic silica beads

Optical microscopy, EDX and STEM images of the obtained beads are shown in **Figure 7**. These images, as well as those shown in **Figures S4, S5, S6 and S7** confirmed the production of microbeads for all the conditions. The observation of the sunflower oil phase after the first centrifugation step as well as the different cleaning solutions does not reveal the presence of iron oxide nanoparticles, suggesting a complete incorporation of the nanoparticles into the silica structure. Microscopy images also suggested a better distribution of the magnetic nanoparticles inside the silica structure when using Span 80. An increase of the bead diameter was observed with increased content of $Fe_3O_4$ nanoparticles as presented in **Table 2**. This is likely to be the result of the space occupied by the magnetic nanoparticles in the structure. Quantitative EDX was also tried to determine the iron oxide loading. Atomic absorption spectroscopy is likely to be a better method and will investigated in the future.

**Table 2.** Summary of the results obtained for the preparation of magnetic silica beads.

| Sample composition | Diameter [µm] (Optical microscopy) | Diameter [µm] (SEM) | Diameter [µm] (TEM) | BET Surface area [m$^2$ g$^{-1}$] | Preparation yield [%] | Zeta potential [a] [mV] |
|---|---|---|---|---|---|---|
| **Samples prepared without Span 80** | | | | | | |
| 0 wt% $Fe_3O_4$ | 2.7 ± 1.3 | 2.8 ± 0.9 | - | - | - | - |
| 3 wt% $Fe_3O_4$ | 3.6 ± 1.2 | 3.7 ± 1.6 | - | - | 23 ± 1 | - 17.9 ± 0.7 |
| 6 wt% $Fe_3O_4$ | 4.1 ± 0.9 | 3.3 ± 1.1 | - | 18.7 | 59 ± 4 | -16.2 ± 0.5 |
| 10 wt% $Fe_3O_4$ | 4.5 ± 0.9 | 3.5 ± 1.2 | 3.9 ± 1.2 | 17.1 | 61 ± 5 | - 18.9 ± 0.6 |
| 15 wt% $Fe_3O_4$ | 4.7 ± 1.0 | 4.0 ± 1.3 | - | 17.3 | 25 ± 2 | - 17.9 ± 0.7 |
| **Samples prepared with 1 wt% Span 80** | | | | | | |
| 0 wt% $Fe_3O_4$ | 2.8 ± 1.2 | 3.8 ± 1.2 | - | - | - | - |
| 3 wt% $Fe_3O_4$ | 5.2 ± 1.4 | 6.2 ± 2.4 | - | - | 30 | - 15.4 ± 0.8 |
| 6 wt% $Fe_3O_4$ | 5.4 ± 1.6 | 3.9 ± 1.4 | - | 1.7 | 61 ± 7 | - 15.1 ± 0.4 |
| 10 wt% $Fe_3O_4$ | 4.9 ± 1.1 | 4.6 ± 1.1 | 4.1 ± 0.9 | 1.0 | 84 ± 8 | -14.8 ± 0.2 |
| 15 wt% $Fe_3O_4$ | 5.2 ± 1.5 | 5.3 ± 1.8 | - | 0.4 | 81 ± 13 | - 13.9 ± 0.6 |
| **Samples prepared with 2 wt% Span 80** | | | | | | |
| 0 wt% $Fe_3O_4$ | 4.8 ± 2.0 | 5.4 ± 1.9 | - | - | - | - |
| 3 wt% $Fe_3O_4$ | 7.8 ± 2.5 | 10.4 ± 6.3 | - | - | 30 ± 6 | - 11.1 ± 0.3 |
| 6 wt% $Fe_3O_4$ | 8.1 ± 2.5 | 6.3 ± 1.7 | - | 11.4 | 52 ± 4 | - 12.3 ± 0.5 |
| 10 wt% $Fe_3O_4$ | 8.6 ± 2.1 | 5.7 ± 2.3 | 6.0 ± 1.7 | 6.7 | 53 ± 15 | - 11.2 ± 0.4 |
| 15 wt% $Fe_3O_4$ | 8.5 ± 2.3 | 6.2 ± 1.7 | - | 15.3 | 52 ± 17 | - 9.9 ± 0.2 |

[a] (Zeta potential of the beads measured at pH 7)

The preparation yields of magnetic silica beads varied between 23 and 84 % depending on the experimental conditions. An increase of the yield was observed for samples prepared with higher concentrations of $Fe_3O_4$ nanoparticles as shown in **Table 2**. This is likely to be due to the higher incorporation of magnetic nanoparticles inside the silica bead structure enabling the beads to be easily recovered from solution during the different cleaning stages. An impact of the Span 80 content on the preparation yield was also observed. Samples prepared with 1 wt% surfactant exhibited yields up to 50% higher than the ones prepared with 2 wt% of surfactant. This could be due to the higher roughness of the beads prepared with 2 wt % resulting in increased inter-bead interactions making their redispersion more difficult during the different cleaning stages.

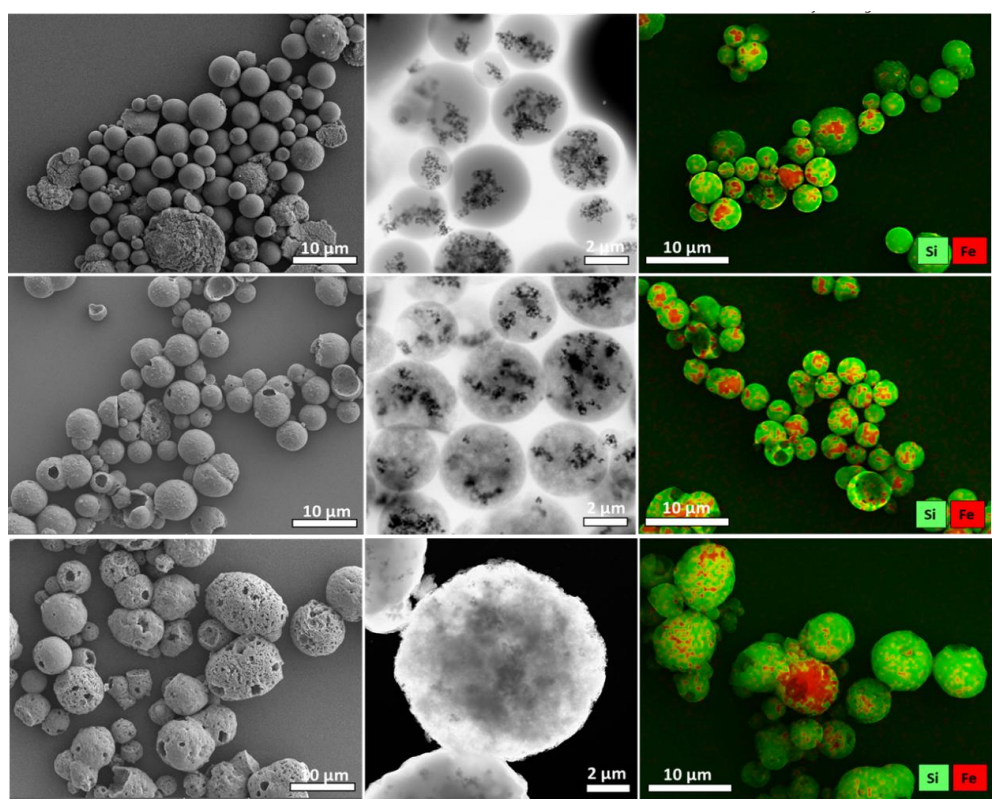

**Figure 7:** Images of magnetic silica beads produced using the Silverson mixer at 4000 RPM for 60 seconds with 30 wt% Ludox HS40 and 10 wt% $Fe_3O_4$ in the aqueous phase and 0 (top row), 1 (middle row) and 2 (bottom row) wt% of Span 80 in the oil phase. Images obtained with SEM (left column), STEM (middle column) and EDX (right column).

SEM and STEM images, presented in **Figure 7,** highlighted a change of porosity induced by addition of Span 80. A homogeneous grey color was observed for the beads prepared without Span 80 suggesting an absence or at least very limited porosity of the structure. In the presence of Span 80, a non-uniform contrast was observed suggesting an internal porosity with a higher level of macroporosity for increased concentrations of surfactant. BET measurements were carried out to quantify the porosity (**Table 2**). Silica particles are rigid and thermally stable, resulting in the formation of nano-size gaps after colloidal destabilization. These generate a surface area larger than the 0.1 - 0.3 $m^2 g^{-1}$ measured for a non-porous sphere of diameter 3 to 8 µm composed of silica and iron oxide. On addition of Span 80, a large decrease of the surface area is measured at low concentration. At higher concentration, an increased porosity is observed.

To rationalize these seemingly contradictory results we note that STEM enables the visualization of the system's macroporosity. BET measurements, being based on the adsorption and desorption of nitrogen, give access to the micro and mesoporosity. As suggested with STEM, increasing the amount of Span 80 generates increased roughness and macroporosity of the system. However, that enhanced macroporosity also tends to remove the nano-channels formed between silica nanoparticles, leading to a decrease of the surface area measured at 1wt% Span 80. With 2 wt% Span 80, the large increase of both roughness and macroporosity of the system balance the disappearance of nano-channels resulting in an increase of the surface area. The surface area measured for the different experimental conditions appear quite low compared to the 300 to 800 $m^2 g^{-1}$ usually obtained for mesoporous magnetic silica nanobeads. [13, 14, 16, 17] However, the beads presented in this paper are 10 to 40 times larger than the nanobeads with a correspondingly smaller surface area.

The tunable roughness and macroporosity are promising for applications requiring high surface area such as catalysis or adsorption. However, it also impacts the strength of the beads. Microbeads produced using less than 1 wt% Span 80 were easily redispersed after vacuum drying. Those beads maintained a spherical structure after drying confirming their strength and stability. However, it was not possible to properly redisperse the magnetic beads produced with 2 wt% Span 80. Large agglomerated structures composed of silica and $Fe_3O_4$ were observed suggesting a weaker structure. The optical microscopy images taken during this test are presented in **Figure S8**.

The magnetic properties of the beads were studied under an external magnetic field. The results in **Figure 8a** show that it is possible to recover the magnetic silica beads from solution in less than 30 seconds using a magnet. This enhanced recovery is promising for applications requiring the adsorption and separation of chemicals. The observation of the samples by optical microscopy after application of a magnet suggested a residual magnetic moment. The beads tend to organize into a wormlike structure after a few seconds as shown in **Figure 8b**. Observations carried out after a few days show that the beads maintain this magnetic moment. However, the structure was easily disrupted by mixing, enabling the redispersion of the beads in solution.

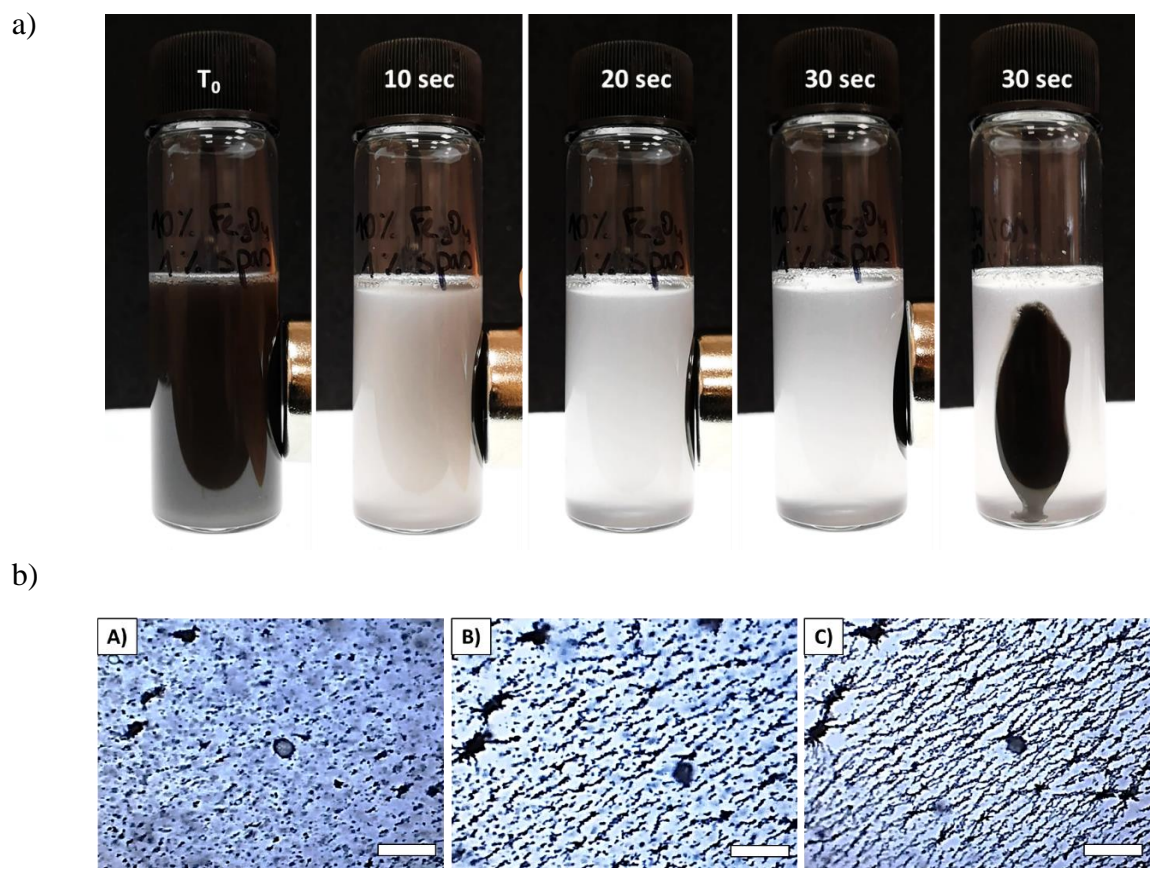

**Figure 8:** a) Images of the separation of magnetic silica beads from the solution under application of an external magnetic field. Sample prepared using the Silverson mixer at 4000 RPM for 60 seconds with 30 wt% Ludox HS40 and 10 wt% $Fe_3O_4$ in the aqueous phase and 1wt % of Span 80 in the oil phase. b) Optical microscopy images of magnetic silica beads after having been exposed once to a magnet. Images taken just after mixing (A), after 7 (B) and 15 (C) seconds without application of any external magnetic field. The picture scale bar is 100 µm.

Zeta potential measurements showed values between -10 and -20 mV suggesting an incipient electrostatic stability. A slight decrease of the values was observed with increased amount of Span 80, likely to be a result of the increased macroporosity of the samples. However, the difference is small. A summary of all the results obtained on the magnetic silica beads is presented in the **Table 2**.

**4. Future work**

The controlled aggregation within emulsion droplets appears to be a promising method to produce composites. Future experiments will investigate the incorporation of additional metals such as $TiO_2$ to obtain a system with photocatalytic properties. Additional research will also be carried out to reduce the diameter of the beads using an ultrasonic homogenizer. The objective being to produce a system with a size below 300 nm enabling a significant increase in the available surface area and enabling their use in medical applications such as medical imaging or targeted drug delivery.

**5. Conclusion**

This work demonstrates the achievability of producing magnetic silica beads via salt-induced destabilization of nanoparticles contained in a reverse emulsion. The use of an emulsion as structural template leads to the formation of silica-based spherical beads as opposed to the fractal structure usually observed from colloidal aggregation. [19] This method enables the safe and fast production of silica beads at room temperature as opposed to conventional multi-step methods based on growing of a silica layer on a magnetic core, which require various organic solvents and high temperatures. [12-18]

This preparation method also exhibits a high level of tunability in term of bead size, morphology and chemical composition. Diameters between 1 and 9 µm are accessible depending of the emulsification method. The addition of an increased concentration of hydrophobic surfactant Span 80 to the oil phase enables the easy modification of the beads' roughness and porosity at room temperature. Finally, iron oxide nanoparticles are easily incorporated to the structure, via dispersion in the aqueous phase, conferring magnetic properties to the final system.

Such a preparation approach appears to be a promising alternative to the Stöber method and could be considered for the production of a new generation of co-assembled metal oxide silica composites, which are safe, stable and easily chemically functionalizable. Two specific aspects will now be investigated: the production via ultrasonic emulsification of sub-micron beads for biological applications (e.g. medical imaging, diagnosis systems, drug delivery, bio-separation …) and the production of multi – metal composite materials (e.g. $TiO_2/Fe_3O_4$ @ $SiO_2$, $ZnO/Fe_3O_4@SiO_2$ …) for potential applications in catalysis and water treatment.

## Supporting Information

Supporting Information is available.

## Acknowledgements

The authors thank Dr Heather Greer (Department of Chemistry), Dr David Madden and Dr. Ceren Camur (Department of Chemical Engineering and Biotechnology) for helping with the imaging characterizations (SEM and TEM) and BET measurements. D.F.F Brossault thanks the University of Cambridge for funding through a WD Armstrong PhD scholarship.

## Conflict of interest

The authors declare no conflict of interest.